# RESPONSE TO THE LETTER TO THE EDITOR

By Clifford Spiegelman, S. J. Sheather, W. A. Tobin,
W. D. James, S. Wexler and D. M. Roundhill

*Texas A&M University, Texas A&M University, Forensic Engineering
International, Texas A&M University, Hightstown High School
and Chem Consulting*

This paper has attracted interest around the world from the media (both TV and newspapers). In addition, we have received letters, emails and telephone calls. One of our favorites was a voicemail message asking us to return a call to Australia at which point we would learn who really killed JFK.

We welcome the opportunity to respond to the letter to the editor from Mr. Fiorentino.

Mr. Fiorentino claims that our "statement relating to the likelihood of a second assassin based on the premise of three or more separate bullets is demonstrably false." In response we would like to simply quote from page 327 of Gerald Posner's book *Case Closed*, one of the most well known works supporting the single assassin theory: "If Connally was hit by another bullet, it had to be fired from a second shooter, since the Warren Commission's own reconstructions showed that Oswald could not have operated the bolt and refired in 1.4 seconds."

Mr. Fiorentino also claims that the "second fatal flaw is the use of a rather uncomplicated formula based on Bayes Theorem." Let $E$ denote the evidence and $T$ denote the theory that there were just two bullets (and hence a single shooter). We used Bayes Theorem to hypothetically calculate $P(T|E)$ from $P(E|T)$ and the prior probability $P(T)$. In order to make $P(T|E)$ ten times more likely than $P(\bar{T}|E)$, the ratio of the prior probabilities [i.e., $P(T)/P(\bar{T})$] would have to be greater than 15. Thus, we again conclude that this casts serious doubt on Dr. Guinn's conclusion that the evidence supported just two bullets. Sadly, this is far from the first time that probability has been misunderstood and/or misapplied in a case of public interest. A notable British example is the Clark case. See Nobles and Schiff (2005) for details.

Finally, we welcome and, in fact, encourage members of the scientific community to provide alternative analyses of the data.

## REFERENCE

NOBLES, R. and SCHIFF, D. (2005). Misleading statistics within criminal trials: The Sally Clark case. *Significance* **2** 17–19. MR2224077

C. SPIEGELMAN  
S. J. SHEATHER  
DEPARTMENT OF STATISTICS  
TEXAS A&M UNIVERSITY  
3143 TAMU  
COLLEGE STATION, TEXAS 77843-3143  
USA  
E-MAIL: cliff@stat.tamu.edu

W. A. TOBIN  
FORENSIC ENGINEERING INTERNATIONAL  
2708 LITTLE GUNSTOCK RD.  
LAKE ANNA, VIRGINIA 23024-8882  
USA

W. D. JAMES  
CENTER FOR CHEMICAL CHARACTERIZATION  
  AND ANALYSIS  
TEXAS A&M UNIVERSITY  
3144 TAMU  
COLLEGE STATION, TEXAS 77843-3144  
USA

S. WEXLER  
HUMANITIES AND ADVANCED  
  PLACEMENT GOVERNMENT  
HIGHTSTOWN HIGH SCHOOL  
25 LESHIN LANE  
HIGHTSTOWN, NEW JERSEY 08520  
USA

D. M. ROUNDHILL  
CHEM CONSULTING  
13325 BLACK CANYON DRIVE  
AUSTIN, TEXAS 78729  
USA